\documentclass[pra,superscriptaddress,nofootinbib,11pt,a4paper]{revtex4-1}
\pdfoutput=1

\usepackage{subfigure}
\usepackage[export]{adjustbox}
\usepackage{amsmath,bm,amssymb}
\usepackage{hyperref}
\usepackage{euscript,array}
\usepackage{wrapfig}
\newcommand{\bea}{\begin{eqnarray}}
\newcommand{\eea}{\end{eqnarray}}
\newcommand{\beq}{\begin{equation}}
\newcommand{\eeq}{\end{equation}}

\newcommand{\bega}{\begin{align}}
\newcommand{\eega}{\end{align}}

\begin{document}

\title{The Casimir effect for nonlinear sigma models and the Mermin-Wagner-Hohenberg-Coleman theorem}

\author{Antonino Flachi}
\affiliation{Department of Physics \& Research and Education Center for Natural Sciences,\\ Keio University, 4-1-1 Hiyoshi, Yokohama, Kanagawa 223-8521, Japan}
\email{flachi@phys-h.keio.ac.jp}

\author{Vincenzo Vitagliano}
\affiliation{Institut de F\'{i}sica d'Altes Energies (IFAE), The Barcelona Institute of Science and Technology (BIST), Campus UAB, 08193 Bellaterra (Barcelona), Spain}
\email{vvitagliano@ifae.es}

\begin{abstract}
{The quantum vacuum (Casimir) energy arising from noninteracting massless quanta is known to induce a long-range force, while decays exponentially for massive fields and separations larger than the inverse mass of the quanta involved. Here, we show that the interplay between dimensionality and nonlinearities in the field theory alters this behaviour in a nontrivial way. We argue that the changes are intimately related to the Mermin-Wagner-Hohenberg-Coleman theorem, and illustrate this situation using a nonlinear sigma model as a working example. We compute the quantum vacuum energy, which consists of the usual Casimir contribution plus a semiclassical contribution, and find that the vacuum-induced force is long-ranged at large distance, while displays a complex behaviour at small separations. Finally, even for this relatively simple set-up, we show that nonlinearities are generally responsible for modulations in the force as a function of the coupling constant and the temperature. }
\end{abstract}

\maketitle

\section{Introduction}
In quantum field theory, a continuous symmetry cannot be spontaneously broken in $D=1$ spatial dimension. This fundamental result, due to Mermin and Wagner, Hohenberg, and Coleman \cite{Mermin:1966, Hohenberg:1968, Coleman:1973}, follows from the argument that when a continuous symmetry is spontaneously broken, massless Goldstone bosons emerge. Since in $D=1$ the correlation function of massless bosons suffers from a pathological (infrared divergent) behaviour, it follows that spontaneous symmetry breaking cannot occur. The same is true in $D=2$ spatial dimensions at finite temperature.
The beauty of this result lies in its indifference to details: \textit{any} mechanism capable of inducing the breakdown of a continuous symmetry gets halted once the theory is restricted to $D=1$ (or $D=2$ at finite temperature), regardless of the complexity of the system or the details of the quantum field theory. This simple observation hides an important repercussion of the Mermin-Wagner-Hohenberg-Coleman (MWHC) theorem for the Casimir effect. 

In its original formulation \cite{Casimir:1948}, the Casimir effect refers to the deformations of the electromagnetic quantum vacuum fluctuations caused by the presence of two flat, parallel, and perfectly conducting plates and to the resulting force attracting the plates towards each other. It was later realised \cite{Fisher:1978} that a key feature to the Casimir effect is the presence of massless quanta that induce long-range correlations, a perspective that pointed at much broader implications (for instance, in quantum liquids or superfluids, where long-range correlations may exist due to Goldstone modes of a broken continuous symmetry; see \cite{Kardar:1997cu,Volovik:2003fe,Schecter:2014} for some examples). 
 
While the attractive nature of the Casimir force can be understood, at least in some cases, as a consequence of a reflection symmetry between the boundaries \cite{Kenneth:2006}, the scaling of the force with the distance is a direct consequence of the conformal invariance, that is the massless-ness of the quanta involved ({\em e.g.}, the electromagnetic field in the case discussed by Casimir). On the other hand, in the presence of a mass $m$, the Casimir force scales, in $D$ spatial dimensions, as $\left|F_c\right| \sim (m\ell)^{D/2}{\exp(-m \ell)/ \ell^{D+1}}\left(1+ O(m \ell)\right)$ (This formula refers to the set-up of two $D-1$ dimensional flat parallel boundaries with periodic boundary conditions; however, the same exponential suppression occurs for other boundary conditions; see Ref.~\cite{Elizalde:1989,Mamayev:1979} for an elementary derivation). Thus, unless the separation is comparable with the Compton wavelength of the massive quanta involved, the Casimir force is exponentially suppressed. This explains why, for example, it is safe to ignore the contribution of electrons (whose Compton wavelength, $\lambda_e$, is of the order of $2.4 \times 10^{-12}$m) in macroscopic applications of the Casimir effect. Likewise, the Casimir effect for any other massive (Standard Model) field is thought to have no implications at distances above the (hundreds of) nanometer range, where the Casimir force is routinely measured (e.g., \cite{Bordag:2009,Klimchitskaya:2009,Decca:2015}).

However, once the quantum field theory under consideration is nonlinear, even if the propagating quanta are effectively massive, the Casimir force need not be exponentially suppressed. It may, in fact, be long-ranged, as in the massless case, or display an even more intricate behaviour, depending on the strength of the interaction or on temperature. We will present our argument explicitly, focusing on a simple example of an $\mathbb{O}(N)$ model (for some basics of these models see, for instance, the textbooks \cite{Zinn-Justin:2002,Shifman:2012}) governed by an action of the form 
\bea
\EuScript S = \int d^Dx dt \left\{ \left| \partial_{\mu} n_i \right|^2 - M^2\left(\left| n_i \right|^{2}- r\right) \right\},
\label{eq1}
\eea 
where the $n_i$ ($i=1,2,\cdots, N$) are complex scalar fields, $M$ is an effective mass, $r$ is a coupling constant, and $D$ represents the spatial dimensionality. Treating the (squared) effective mass $M^2$ as an auxiliary field ({\em viz.} as a Lagrange multiplier), the theory is equivalent to the standard nonlinear sigma model \cite{Abbott} with the usual constraint on the norm of the vector $\left| n_i \right|^{2}=r$. In the limit $r\to 0$, the above action describes $N$ free complex scalar fields with mass $M$, for which the Casimir energy per degree of freedom, $\mathcal{E}_{\textrm{Cas}} \equiv {E}_{\textrm{Cas}} /2 N$ (assuming periodic boundary conditions\footnote{This set-up is different from that of two disconnected boundaries (eg, the parallel plates of Casimir's set-up). For periodic boundary conditions there are no boundaries, but there is still a Casimir force that makes the circle shrink or expand, depending on the sign of the force (For an elementary discussion, see the section `The scalar Casimir effect on the circle' of Ref.~\cite{Bordag:2009} pp. 24-26).}) can be written as 
\bea
\mathcal{E}_{\textrm{Cas}} = -  {1 \over \pi\ell}  \int_{M \ell}^\infty {\sqrt{y^2 - M^2\ell^2} \over e^y-1} dy.
\label{eq2}
\eea 
The zero mass limit returns $\mathcal{E}^{(0)}_C = -   {\pi / 6\ell}$, while the ``large mass'' limit, $M\ell \gg 1$, gives 
$\mathcal{E}_{\textrm{Cas}} \sim  \mathcal{E}^{(0)}_C  \sqrt{M \ell} e^{-M\ell}$, that clearly shows the exponential suppression induced by the mass (see Refs.\cite{Bordag:2009,Mamayev:1979}).

In the present case, with a non-vanishing coupling constant, $r$, and fields forced to obey a constraint (i.e., in the presence of field nonlinearities), the effective mass $M^2$ is no longer a free parameter, rather it is fixed according to the constraint imposed on the fields (i.e., by the gap equation) and \emph{not by hand}. If we fix the dimensionality to be $D=1$, then the verdict of the MWHC theorem is final: no transition to a massless phase can occur and $M\neq 0$ for any value of the separation $\ell$. This implies that quantum fluctuations are effectively massive and, according to the above discussion, the Casimir force arising from such fluctuations should decay exponentially for separations larger than the inverse mass-gap, that is in the regime $M \ell \gg 1$. However, this view would be intuitive, but naive: in the present situation (differently from the case of free scalar fields), the mass suppression appearing in the Casimir force is dictated by the gap equation (that determines how $M^2$ depends on the size of the system, $\ell$, or any other external forcing eventually present), inducing in the Casimir force an additional nonlinear dependence on the separation. It is this implication of the MWHC theorem that causes a dependence of the effective mass $M$ on the separation and modifies the exponential behaviour in the Casimir force. This behaviour is already evident since, at least, the seminal work of Ref.~\cite{Luscher:1980ac} where the Casimir energy term has been shown to have a dependence scaling as the inverse size of the system, and here, we clarify this by extending the analysis to higher dimensions. Furthermore, in the following, we numerically calculate the total quantum vacuum force and show that it consists of the usual Casimir term, analogous in 1 spatial dimension to that of Ref.~\cite{Luscher:1980ac} plus a contribution proportional to $M^2$ (this second contribution is of semiclassical nature, since $M^2$ is determined by the one-loop effective equations) and show that, despite the relative simplicity of the set-up, the resulting quantum vacuum force displays a nontrivial behaviour.

\section{Dimensionality and mass gap} 
 The vacuum of the classical theory \eqref{eq1} is degenerate: anyone of the points of the $S^{N-1}$ sphere of radius $\sqrt{r}$ is a valid ground state. Once that one of the vacua is picked up, the original $\mathbb{O}(N)$ symmetry breaks down to $\mathbb{O}(N-1)$ (the symmetry which now leaves the chosen vacuum unchanged), with Goldstone theorem anticipating the occurrence of $(N-1)$ massless bosons. However, the {\em quantum} ground state has here a few different features. The symmetries of the vacuum are determined by the one-loop effective potential: if this potential is extremised by a non-vanishing value of the auxiliary field, say $\widetilde M^2$, then one can expand $M^2$ around  $\widetilde M^2$, giving rise to massive terms for the fields $n_i$ in the action \eqref{eq1}. This disordered phase does not break the $\mathbb{O}(N)$ symmetry, so, rather than Goldstone bosons, the theory contains an $n$-plet of mass $\widetilde M$ particles. However, depending on the dimensionality of the system,  $\widetilde M$ could in principle vanishes at some critical values of size and temperature. If this occurs, it indicates that a symmetry has been broken and the system has experienced a phase transition \cite{Senechal:1993}.
 
The action \eqref{eq1} is bilinear in the fields $n_i$. A Gaussian integration of the fields straightforwardly reveals the following \textit{euclideanised} one-loop effective action at large $N$:
\begin{align}
\EuScript  S^{E}_{\mbox{\tiny{eff}}} &=
 (N-1) \mbox{Tr}\; \log \left(-\Delta - {\partial^2\over \partial \tau^2} + M^2\right)-\int_0^{\beta} d \tau \int d^D x 
\cdot  M^2 \cdot r.
\label{eq4}
\end{align}
The characterization of the mass gap for the ${\mathbb C}P^{N-1}$ and $\mathbb{O}(N)$ models in $D=1$ and subjected to boundary conditions has been discussed extensively (see, for some examples, Refs.~\cite{Flachi:2019jus,Monin,Gorsky,Gorsky2,Bolognesi,Ishikawa:2020eht} and the bibliographies given there).
Here, we are assuming that, besides the compactified Euclidean time, there is only one constrained spatial direction $x_1$, leaving the remaining $D-1$ directions ($x_2, \cdots, x_D$) unconstrained. In practice, we enclose the system within a box of size $\ell_k$ along the direction $x_k$, impose periodic identification, and take the limit $\ell_k\to \infty$ for $k=2,\cdots, D$, leaving the direction $x_1$ confined. For notational convenience we relabel $\ell_1= \ell$ and define $\mbox{V}_D = \left(\prod_{j=1}^D \ell_j \right)$. This is the typical Casimir enclosure. 
Zeta-regularization allows to express the one-loop effective action as (see refs.~\cite{Avramidi,Kirsten,elizalde94,Toms:2012bra})
\begin{align}
\EuScript  S^{E}_{\mbox{\tiny{eff}}} &= 
- (N-1) \left(\zeta(0) \log\Lambda^2 + \zeta' (0)\right)-\int_0^{\beta} d \tau \int d^D x \cdot  M^2 \cdot r,
\label{eq5}
\end{align}
with
\bea
\zeta(s) &=& 
{\mbox{V}_D \over \ell}
\sum_k \sum_{n=-\infty}^\infty \int{d^{D-1}q \over (2\pi)^{D-1}} \left({\bf q}^2 + {p}_k^2 + 4\pi^2 n^2 /\beta^2\right)^{-s}\,,\label{eq6}
\eea
where $\Lambda$ is a normalization constant and the eigenvalues ${p}_k$ are defined by 
\bea
\left({\partial^2 \over \partial x_1^2} + M^2\right) f_k = {p}_k^2 f_k.~~~
\label{eq7}
\eea
Integrating over $q$ in (\ref{eq6}) yields
\bea
\zeta(s) &=& 
{\mbox{V}_D \over (4\pi)^{D-1\over 2} \ell}
{
\Gamma\left({1-D\over 2}+s\right)
\over
\Gamma\left({s}\right)
}
\sum_{k} \sum_{n=-\infty}^\infty \left({p}_k^2 + {4\pi^2 n^2 \over \beta^2}\right)^{{D-1\over 2}-s}.~~~~~~~~~
\label{appeq1}
\eea
Using the Mellin transform, 
\bea   
\lambda^{-z} \Gamma(z) = \int_0^\infty t^{z-1} e^{-\lambda t} dt,
\label{eq8}
\eea
in Eq.~(\ref{appeq1}), it takes only simple steps to arrive at the following representation
\bea
\zeta(s)&=& 
{\mbox{V}_D \over (4\pi)^{D-1\over 2} \ell}
{
1\over
\Gamma\left({s}\right)
}
\int_0^\infty
K(t) \times \Theta(t) 
{dt \over t^{1+{D-1\over 2}-s}}\,,
\label{eq9}
\eea
with $K(t) = \sum_{k} e^{-t p_k^2 }$ being the integrated heat-kernel associated with the differential operator in (\ref{eq7}) and 
\bea
\Theta(t) \equiv \sum_{n=-\infty}^\infty  e^{-4\pi^2 n^2t/\beta^2}
= {\beta \over \sqrt{4 \pi t}} 
\left[1 + 2 \sum_{n=1}^\infty \left(e^{-{\beta^2 n^2 \over 4 t}}\right)\right].
\label{eq10}
\eea
We now write the heat-kernel as follows
\bea
K(t) =
{\ell \over \sqrt{4 \pi t}}{e^{- t M^2 } } \left(1 + \delta K(t)\right), 
\label{eq11}
\eea
where, for periodic boundary conditions, we have
\bea
\delta K(t) =
2 \sum_{k=1}^\infty e^{-{\ell^2 k^2 \over 4 t}}.
\label{eq11bis}
\eea
For different boundary conditions, the decomposition (\ref{eq11}) still holds with a different expression for (\ref{eq11bis}). 
Substituting and performing the integrals over $t$,
we arrive at the following formula
\begin{align}
&\zeta(s)= {\beta \mbox{V}_D}{1\over (4\pi)^{D+1\over 2}}
{\Gamma\left(s-{D+1\over 2}\right) \over \Gamma\left(s\right)}
\left(M^2\right)^{{D+1\over 2}-s}\times\nonumber\\
&
\times\left\{
1
+
{2^{{D+1\over 2}+2-s}\over \Gamma\left(s-{D+1\over 2}\right)} \sum_{n=1}^\infty\left[\left(n \beta M\right)^{\left(s-{D+1\over 2}\right)} K_{{D+1\over 2}-s}\left(n \beta M\right)+\left(n \ell M\right)^{\left(s-{D+1\over 2}\right)} K_{{D+1\over 2}-s}\left(n \ell M\right)\right]\right.\nonumber\\
&
\left.+
{2^{{D+1\over 2}+3-s} \over \Gamma\left(s-{D+1\over 2}\right)}
\sum_{n=1}^\infty \sum_{k=1}^\infty\left(M \sqrt{k^2\ell^2+n^2\beta^2} \right)^{\left(s-{D+1\over 2}\right)} 
K_{{D+1\over 2}-s}\left(M \sqrt{k^2\ell^2+n^2\beta^2} \right)
\right\}\,;
\label{eq12}
\end{align}
here $K_\nu(z)$ is the modified Bessel function of the second kind of order $\nu$. For the derivative we have
\bea
{\zeta'(0)\over {\beta \mbox{V}_D}} &=& 
{1 \over (4\pi)^{D+1\over 2}} \left[
\frac{d}{ds}\left.\frac{M^{D+1-2s}~ {\Gamma\left(s-{{D+1}\over 2}\right)}}{\Gamma(s)}\right|_{s=0}
-
\hat\varpi_D (\beta,0) - \hat\varpi_D (0,\ell)+ \hat\varpi_D (\ell,\beta)\right],\nonumber\\
\label{eq13}
\eea
where
\begin{align}
\hat\varpi_D (x,y) &=&
M^{D+1}~ 2^{{D+1\over 2}+3} \sum_{n,k=1}^\infty \left(M\sqrt{k^2 x^2+n^2y^2} \right)^{-{D+1\over 2}}  K_{{D+1\over 2}}\left(M \sqrt{k^2x^2+n^2y^2} \right).
\label{eq14}
\end{align}
Substituting \eqref{eq12}, (\ref{eq13}) and (\ref{eq14}) in (\ref{eq5}), the one-loop effective action follows at ease.
\subsection{The case $D=1$: Casimir effect on a closed string revisited}\label{secc}
Setting $D=1$ (this is the case where the MWHC theorem implies a non-vanishing mass gap), gives
\begin{align}
\EuScript  S^{E}_{\mbox{\tiny{eff},D=1}} =& 
{\beta }\int d x 
\left\{ - {N - 1 \over 4 \pi }  \left[
M^2\cdot(\hat r-1)+M^2 \log \left({M^2\over \Lambda^2}\right)
 - \varpi_1 (0,\ell) -\varpi_1 (\beta,0)+ \varpi_1 (\beta,\ell) 
\right]\right\},
\label{eq15}
\end{align}
where we have rescaled the coupling, $\hat r=4 \pi r/(N-1)$.
Implementing the constraint $\delta \EuScript  S^{E}_{\mbox{\tiny{eff}}}/\delta M^2 =0$ yields
\begin{align}
\hat r+\log \left(M^2/ \Lambda^2\right) - {\partial  \varpi_1 (0,\ell) \over \partial M^2}- {\partial \varpi_1 (\beta,0) \over \partial M^2}+ {\partial \varpi_1 (\beta,\ell) \over \partial M^2}=0\,.
\label{eq18}
\end{align}
\begin{figure*}[t!]
$\begin{array}{cc}
\subfigure{
\includegraphics[height=4.7cm,valign=t]{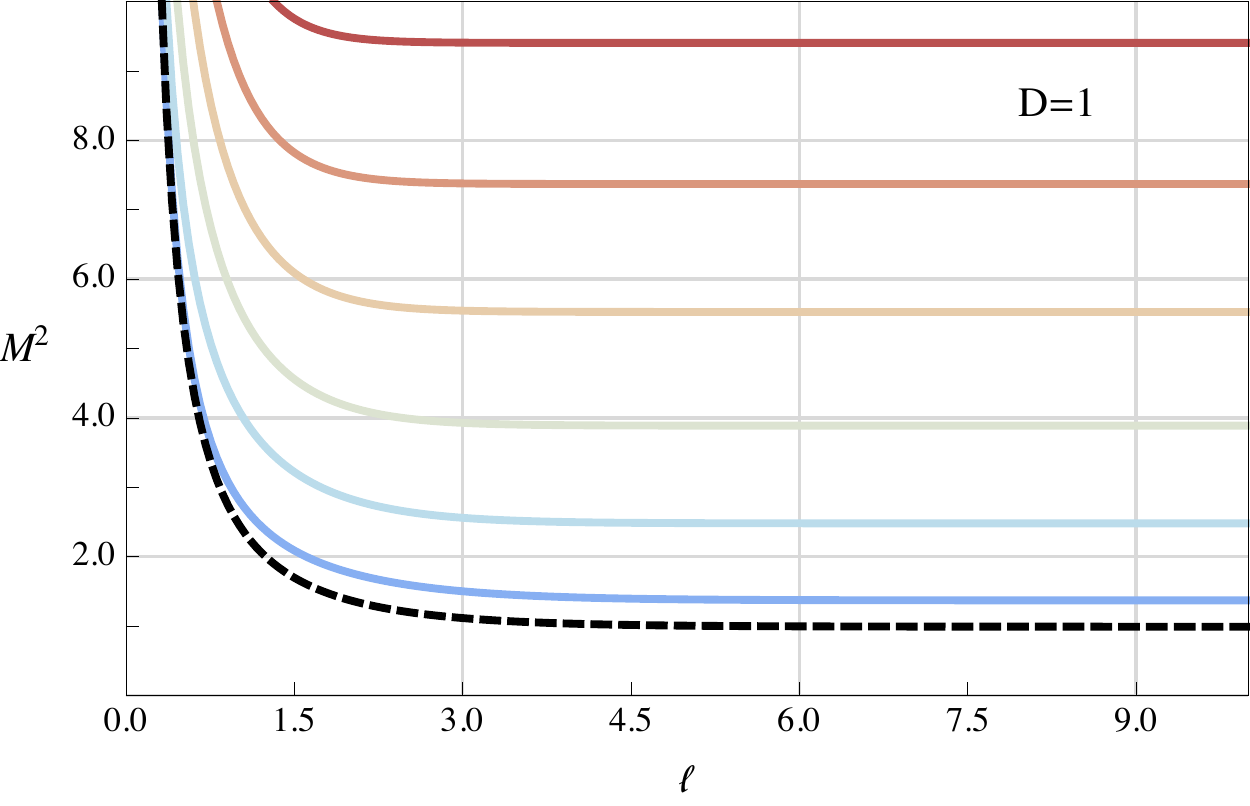}}
\hspace{0.1cm}\subfigure{
\includegraphics[height=4.7cm,valign=t]{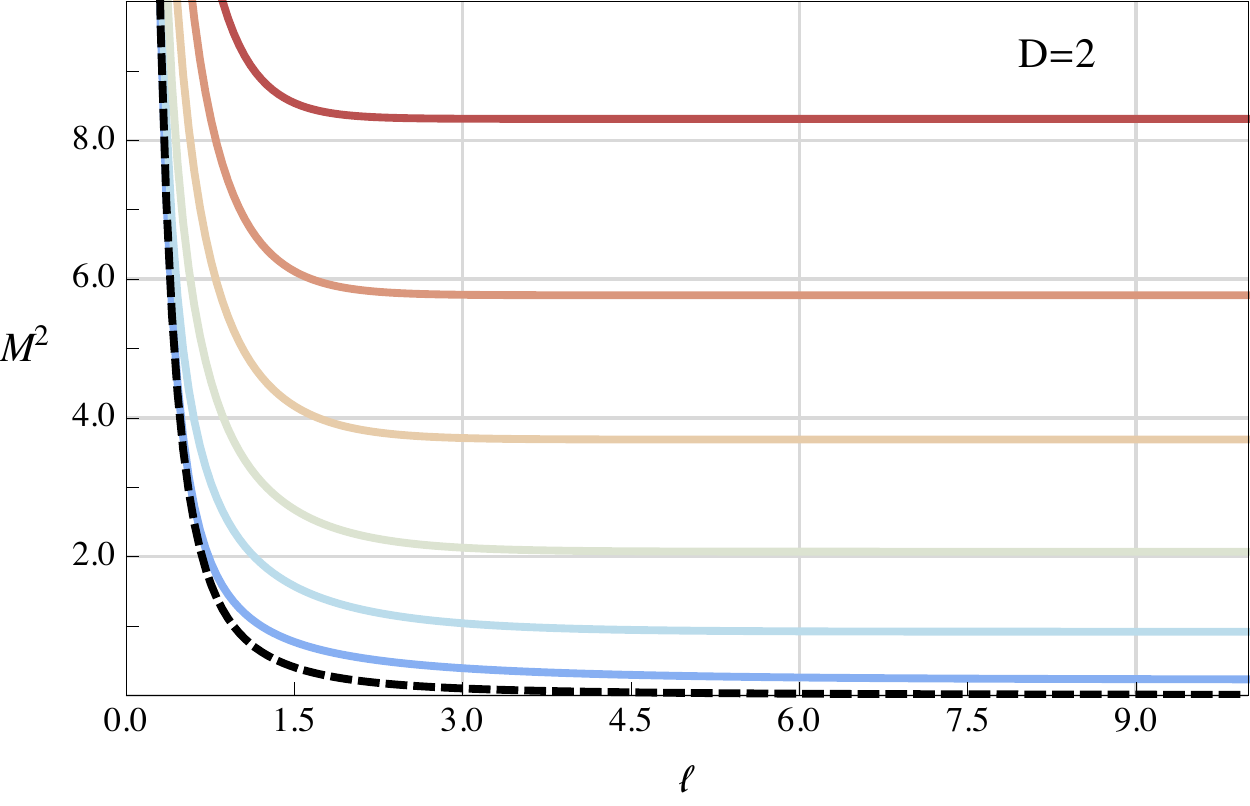}}
\end{array}$\\
\caption{Numerical solution for  $M^2$ in $D=1$ ({\em Left Panel}) and in $D=2$ ({\em Right Panel}) as a function of size $\ell$ and temperature $T$ (we have set  $\hat r=0.01$).
Temperature varies between $T=0.1$ (bottom-blue curve) and $T=10$ (top-red curve). The dashed black curve corresponds to the zero temperature limit.
According to the MWHC theorem, no minimum is allowed in $M^2=0$  for $D=1$ and $T=0$.
In  $D=2$, as expected from the MWHC theorem, the gap equation can be minimised by $M^2=0$  when $T=0$ and $\ell\rightarrow\infty$. For $T\neq0$, however,  $M^2=0$ is not allowed for any $\ell$.
}
\label{fig_1}
\end{figure*}

The equation above, at any given temperature and size, can exhibit different either one or zero roots, depending on the value of the renormalised coupling constant, $\hat r$. The numerical solution  $M^2$ of \eqref{eq18}, and how this changes with $T$ and $\ell$,  is shown in Fig.~\ref{fig_1} (Left Panel). In the zero temperature limit, $\beta\to \infty$, the last two terms in (\ref{eq18}) vanish, as it is easily seen by noticing that $K_p\left(z\right) \sim \sqrt{\pi/ (2z)} e^{-z}$, leaving 
\bea
\hat{r}+\log \left(M^2/\Lambda^2 \right) - {\partial  \varpi_1(0,\ell) \over \partial M^2}=0\,.
\label{eq19}
\eea

In the limit of large $\ell$, the last term in \eqref{eq19} can also be ignored, and, in accordance with the MWHC theorem, the logarithm prevents any minima from occurring at $M^2=0$. A more relevant regime is that of small $\ell$. The small $\ell$ expansion of the Bessel series \cite{Fucci:2014mya} contained in the function $\varpi_1(0,\ell)$ yields
\begin{align}
-\varpi_1(0,\ell) &\approx& {4\pi^2\over 3 \ell^2} - 4 \pi \sqrt{M^2\over \ell^2}-
M^2 \left(2 \gamma_e -1\right) - M^2 \log\left( {\ell^2 M^2\over 16 ~\pi^2}\right) -{1\over 2}{\ell^2 M^4}\zeta'(-2).
\label{eq20}
\end{align}
In this limit the logarithmic singularity in (\ref{eq19}) cancels, but a new non-analytic term (leading to a singularity in the mass gap equation) proportional to $\sqrt{M^2}$ appears, again impeding the mass to attain a zero value. The above results can also be confronted with the results of Ref.~\cite{Bolognesi}.

All these results are trivially extended (by use of the modular symmetry $\ell \leftrightarrow \beta$) to the limit large $\ell$ and high temperature.
Worth of notice is the first term in (\ref{eq20}) that does not contribute to the gap equation; in fact, in the euclidean effective action, this is nothing but the vacuum energy. Multiplying by the overall factor $-(N-1)/4\pi$ and integrating over the separation, one obtains the Casimir energy of the string $E_s =- (N-1) \times {\pi\over 3 \ell}$, that is the Casimir energy of $N-1$ complex scalar fields with periodic boundary conditions, reproducing known results \cite{Luscher:1980ac,Shifman:2007}.

The renormalised mean quantum vacuum energy density can be expressed as usual as the sum over the energies $\omega_n (M^2)$ of the mode fluctuations (or as the integral over the energy density ${E}_{\textrm{vac}}  = \int dx \langle T^{00}(x)\rangle$),
\bea
E_{\textrm{vac}}  =\sum_n \omega_n (M^2) - E_{\infty}\,.
\label{eq21}
\eea
Importantly, in the present case, the energies depend on the mass gap $M^2$ and indirectly on the size of the system.
The quantity $E_{\infty}$ is added to normalize the energy to zero once the infinite volume limit is taken. With respect to the usual noninteracting  situation -- that is $r\to 0$ in \eqref{eq1} --, the present case presents two essential differences. One is that the quantity $M^2$ is not set \textit{a priori}, but is determined by extremisation of the effective action; secondly, the term proportional to $r \times M^2$ in the action also contributes to the vacuum energy by an amount proportional to the mass gap determined at one-loop (this term is a semiclassical contribution, vanishing in the limit $r\to 0$), implying that the vacuum energy has to be computed self-consistently. We implement the calculation in a two-step numerical procedure: we first extremise the effective action and determine the quantity $M^2$, and then compute the Casimir energy according to (\ref{eq21}). The quantity $E_\infty$ is the counter-term that cancels the infinite asymptotic contribution to the energy and is also evaluated numerically, with its value extracted from the non-renormalised energy at fixed and large separation (this process is repeated from increasing values of the separation until the value of $E_\infty$ converges).

\begin{figure*}[t!]
$\begin{array}{cc}
\subfigure{
\includegraphics[height=4.5cm,valign=t]{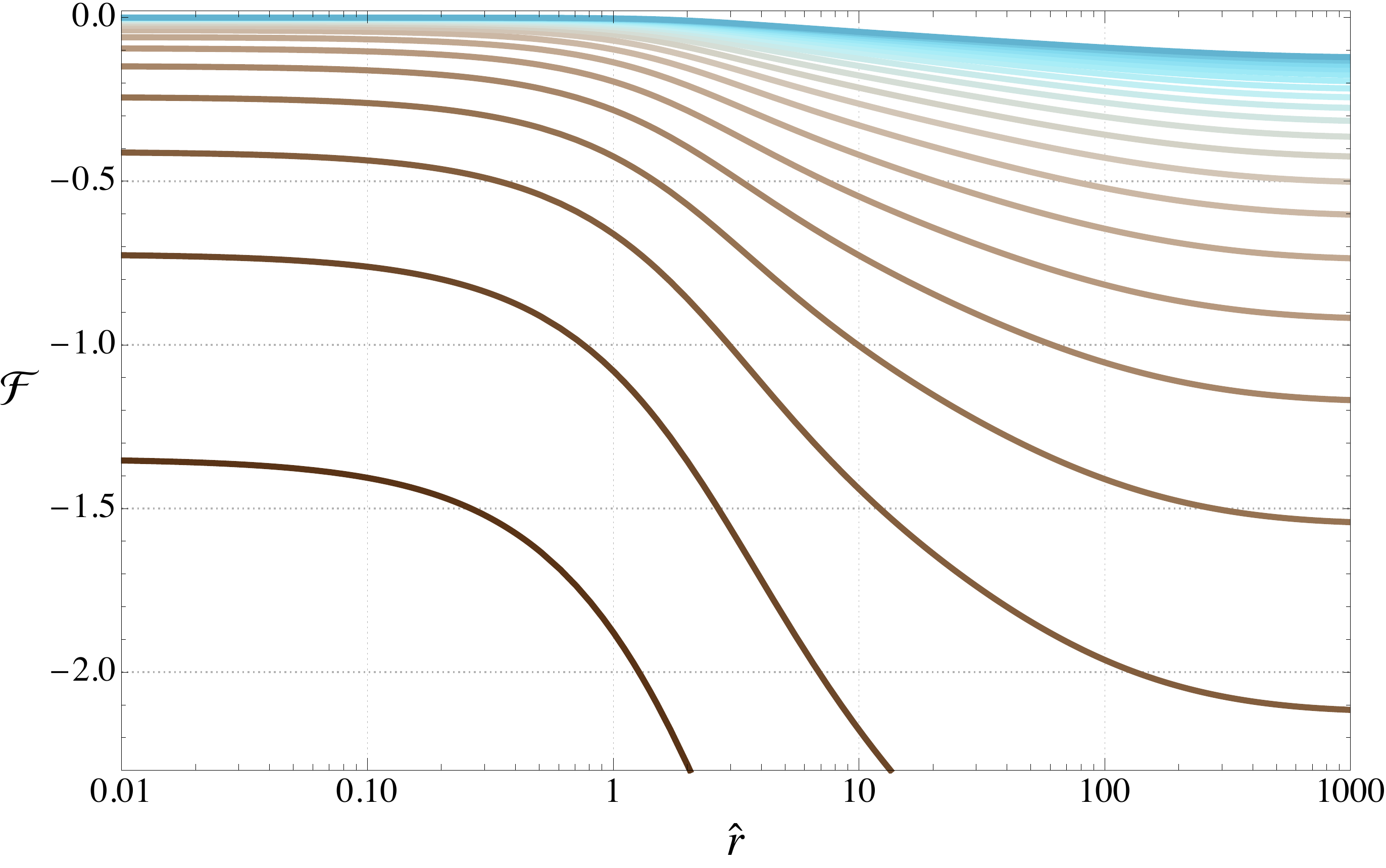}}
\subfigure{
\includegraphics[height=4.5cm,valign=t]{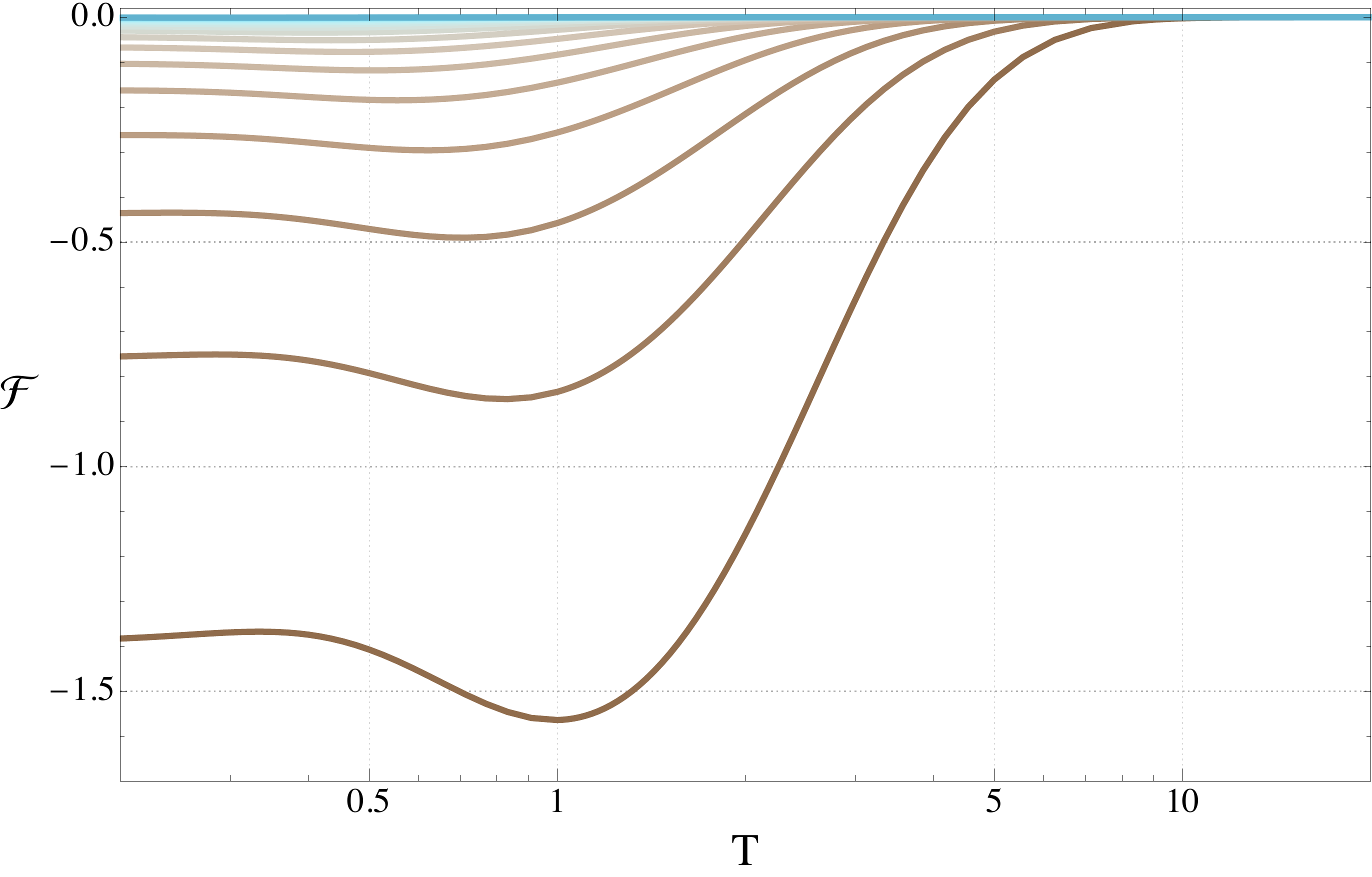}}
\hspace{0.2cm}
\subfigure{
\includegraphics[height=4cm,valign=t]{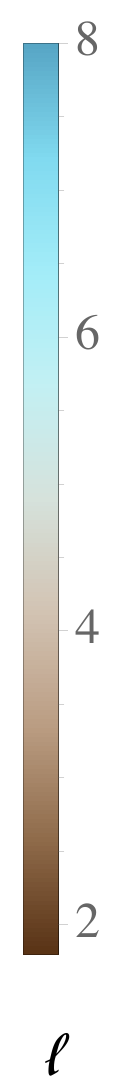}}
\end{array}$\\
\caption{{Evolution of the force along with the variation of the rescaled coupling $\hat{r}$ ({\em Left Panel}, evaluated at $T=0.1$) and of the temperature $T$ ({\em Right Panel}, at $\hat{r}=0.25$), at fixed size $\ell$ and $D=1$. The colour gradient corresponds to different sizes $\ell$, from $\ell=1.8$ (brown) to $\ell=8$ (cyan). Contributions to the free Casimir energy from the semiclassical piece and the temperature reflect in the nontrivial modulation of the corresponding force.}}
\label{fig2}
\end{figure*}

Numerical results are shown in Fig.~\ref{fig2}. For clarity of illustration, we plot the total force per degree of freedom\footnote{In our notation, $\mathcal{E}$ is the energy per degree of freedom.}, $\mathcal{F}_{\textrm{tot}}=-\partial \mathcal{E}_{\textrm{tot}}/\partial \ell$ for two illustrative choices of parameters. The main feature that here arises is that, due to field nonlinearities, at fixed size $\ell$ the Casimir force can be non trivially modulated by an external change of the coupling constant $\hat r$ and the temperature $T$. In both cases, the force initially becomes stronger (more attractive), then decreases in modulus, eventually approaching zero for large values of $T$, or a constant value proportional to the derivative of the L\"uscher term for large $\hat r$. At larger distances (along the color gradient in the figures), the force reduces, as one might expect.

\subsection{The case $D=2$: Casimir effect on a cylinder}

For $D=2$ and $T\neq 0$, the Euclidean effective action for the $\mathbb{O}(N)$ model \eqref{eq1} reads
\begin{align}
&\EuScript S_{\mbox{\tiny{eff},D=2}}^E = \beta \int d^2x\Bigg\{ 
-\frac{(N-1)}{\pi}\Bigg[\frac{  M^2 \hat{r} }{4  }+\frac{  M^3 }{6 }+M ~\left( \frac{\text{Li}_2\left(e^{-M \ell }\right)}{\ell^2}+\frac{ \text{Li}_2\left(e^{-M \beta }\right)}{\beta^2}\right)+
\nonumber\\
&+\frac{\text{Li}_3\left(e^{-M \ell }\right)}{\ell ^3}+\frac{\text{Li}_3\left(e^{-M \beta }\right)}{\beta ^3}+\sum_{k,n=1}^{\infty}\left(\frac{2   M  e^{-M \sqrt{k^2 \ell ^2+\beta ^2 n^2}}}{  \left(k^2 \ell ^2+\beta ^2 n^2\right)}+\frac{2   e^{-M \sqrt{k^2 \ell ^2+\beta ^2 n^2}}}{  \left(k^2 \ell ^2+\beta ^2 n^2\right)^{3/2}}\right)\Bigg]\Bigg\}\,,\label{220}
\end{align}
where $ \text{Li}_\nu(z)$ is the de Jonqui\`{e}re's ({\em aka} polylogarithm, {\em aka} Bose's) function of order $\nu$. In the large $\ell$ limit, both the last term and $ \text{Li}_\nu\left(e^{-M \ell}\right)$ can be neglected. Imposing then the usual constraint $\delta \EuScript  S^{E}_{\mbox{\tiny{eff}}}/\delta M^2 =0$ leads to the gap equation 
\bega
\frac{2}{\beta} \log \left(1-e^{-M\beta}\right)+M+\hat r=0
\,,
\end{align}
which clearly shows the log-singular contribution coming from the finite $T$ part (first term). Again, long range interactions are prevented in agreement with the MWHC theorem. In the zero temperature regime, the logarithmic term disappears as expected: massless phases are allowed at large $\ell$ and $T=0$ (see also Fig.~\ref{fig_1}, {\em Right Panel}). Similarly, due to the compactification of one of the spatial dimensions, finite size effects in the zero temperature limit induce a singular term which also does not allow transitions to a massless phase. 
\begin{figure*}[b!]
$\begin{array}{cc}
\subfigure{
\includegraphics[height=4.5cm,valign=t]{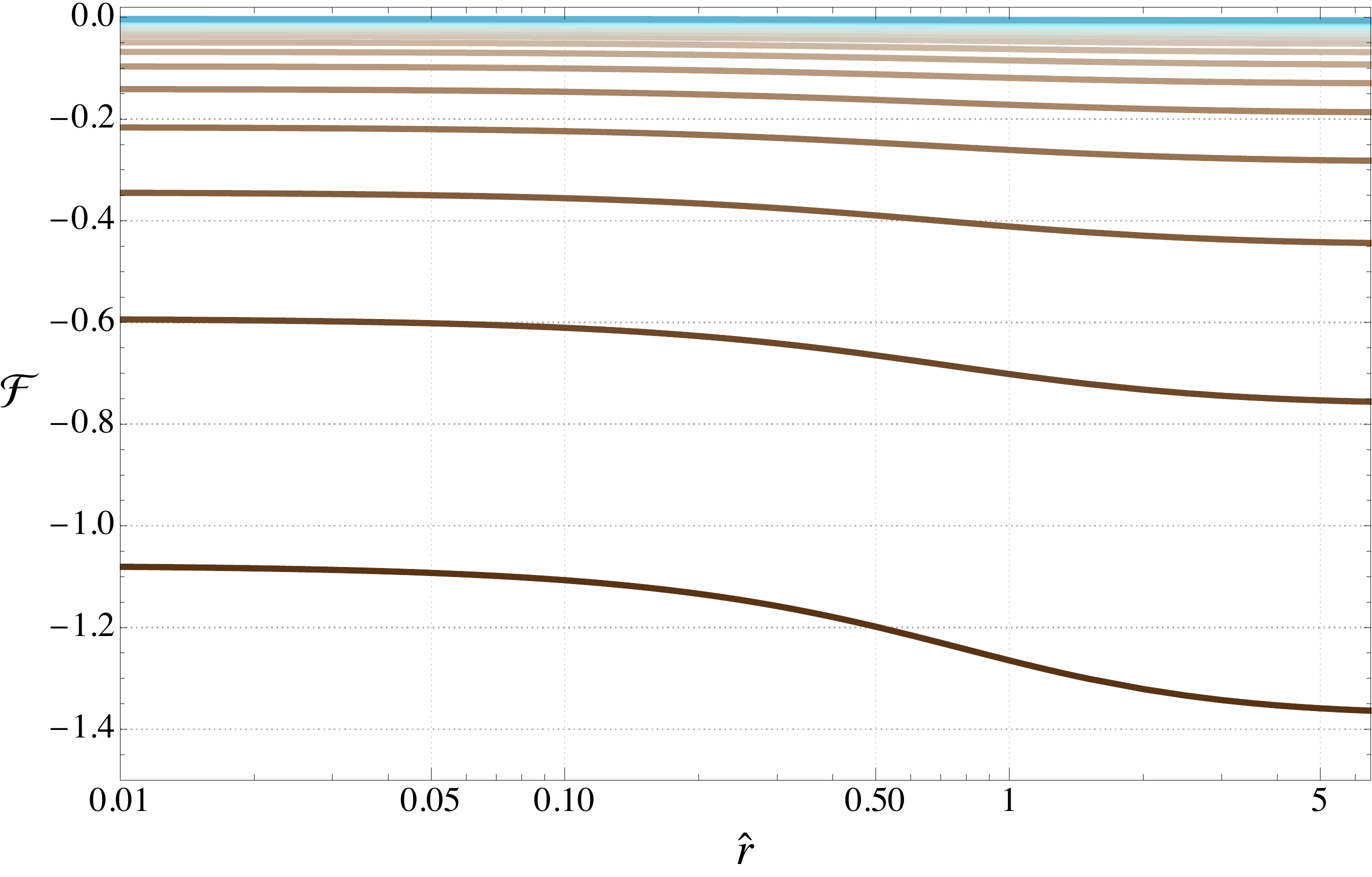}}
\subfigure{
\includegraphics[height=4.5cm,valign=t]{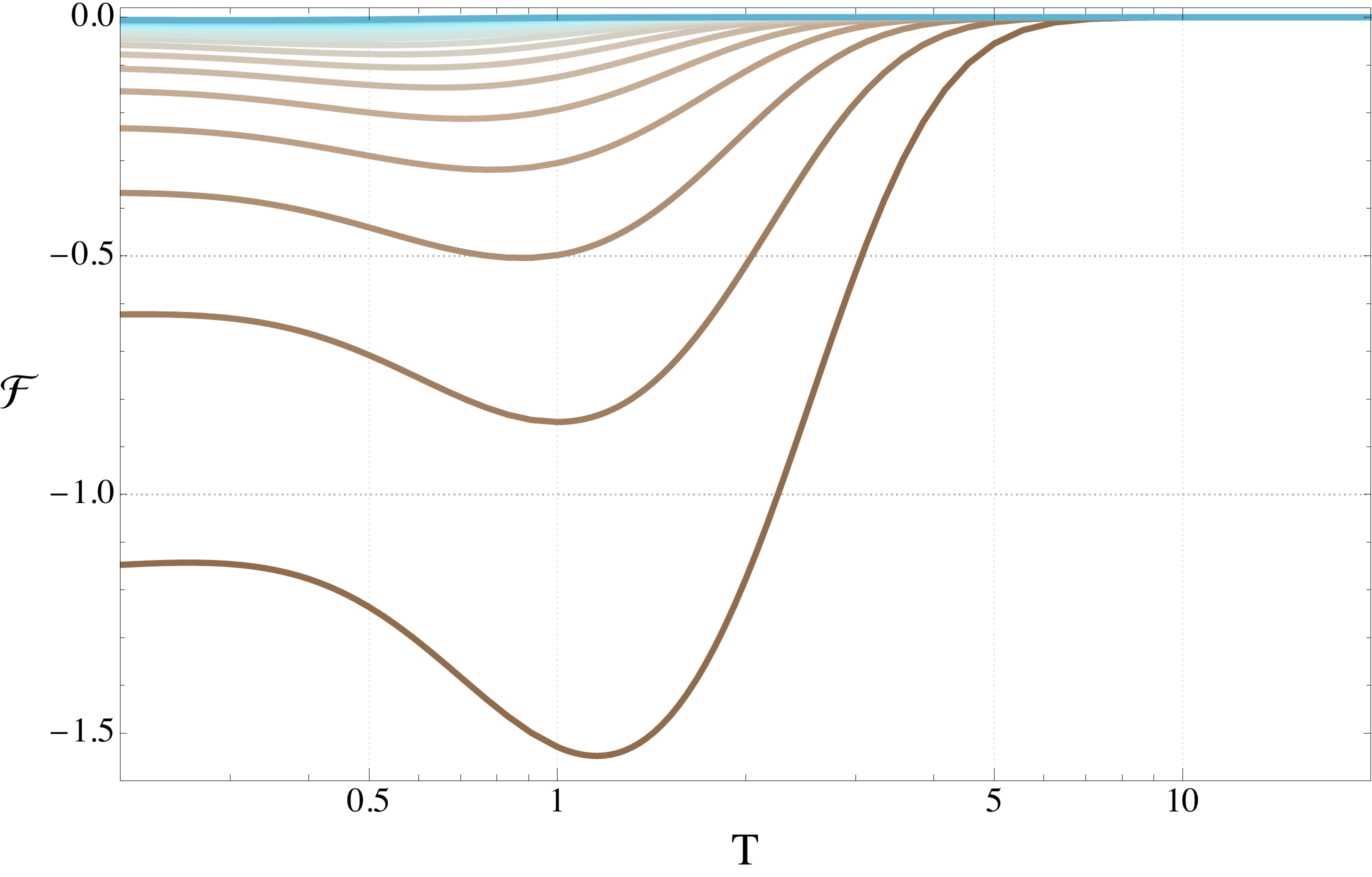}}
\hspace{0.2cm}
\subfigure{
\includegraphics[height=4cm,valign=t]{Lungh.pdf}}
\end{array}$\\
\caption{{Evolution of the force along with the variation of the rescaled coupling $\hat{r}$ ({\em Left Panel}, evaluated at $T=0.1$) and of the temperature $T$ ({\em Right Panel}, at $\hat{r}=0.25$), at fixed size $\ell$ and $D=2$. The color gradient corresponds to different sizes $\ell$, from $\ell=1.8$ (brown) to $\ell=8$ (cyan).}}
\label{fig4}
\end{figure*}

The modulation of the Casimir force as a function of the coupling $\hat r$ and of the temperature $T$, at fixed length, is shown in Fig.~\ref{fig4}. In particular, the {\em Left Panel} shows that, by increasing $\hat r$, the force initially becomes slightly more attractive, and then reaches a constant value which is determined by the small $M$ limit of the $ \text{Li}_3(z)$ terms. To larger values of $\hat r$ correspond in fact smaller values of the mass gap. The last two terms of the first line in \eqref{220} becomes negligible, with $ \text{Li}_3(z)$ becoming the dominant terms. Finally, the {\em Right Panel} describes changes with temperature. The high temperature behaviour is determined by the fact that, at fixed size, higher temperatures correspond to larger values of $M_\ell$, but also to smaller differences between the two values of $M_\ell$ and $M_\infty$. This implies both the (regularised) energy and its derivative to vanish asymptotically.

\subsection{The case $D>2$}

In the higher dimensional case, the MWHC theorem does not forbid a gapless phase: setting $D>2$ yields a log term in the effective action, as in (\ref{eq15}), but multiplied by a $M^{1+D}$ factor that effectively removes the singularity (see (\ref{eq12})-(\ref{eq14})). For sake of clarity, we set $D=3$. Then,
\begin{align}
\EuScript S_{\mbox{\tiny{eff},D=3}}^E =- \beta \int d^3x &
\frac{(N-1)}{4\pi}\left[M^2 \hat r-\frac{ M^4}{8 \pi }\log {  M^2 \over \Lambda^2 }+\frac{3 M^4}{16 \pi }+
\frac{\varpi_3 (\beta,\ell) -\varpi_3 (\beta,0)-\varpi_3(0,\ell)}{4 \pi}\right]\,,
\end{align}
and the gap equation
\begin{align}
M^2- M^2 \log \frac{M^2}{\Lambda^2}+4 \pi\hat r+\frac{\partial}{\partial M^2}\left[\varpi_3 (\beta,\ell) -\varpi_3 (\beta,0)-\varpi_3(0,\ell)\right]=0
\,.\label{gapd3}
\end{align}

\begin{figure*}[b!]
$\begin{array}{cc}
\subfigure{
\includegraphics[height=4.7cm,valign=t]{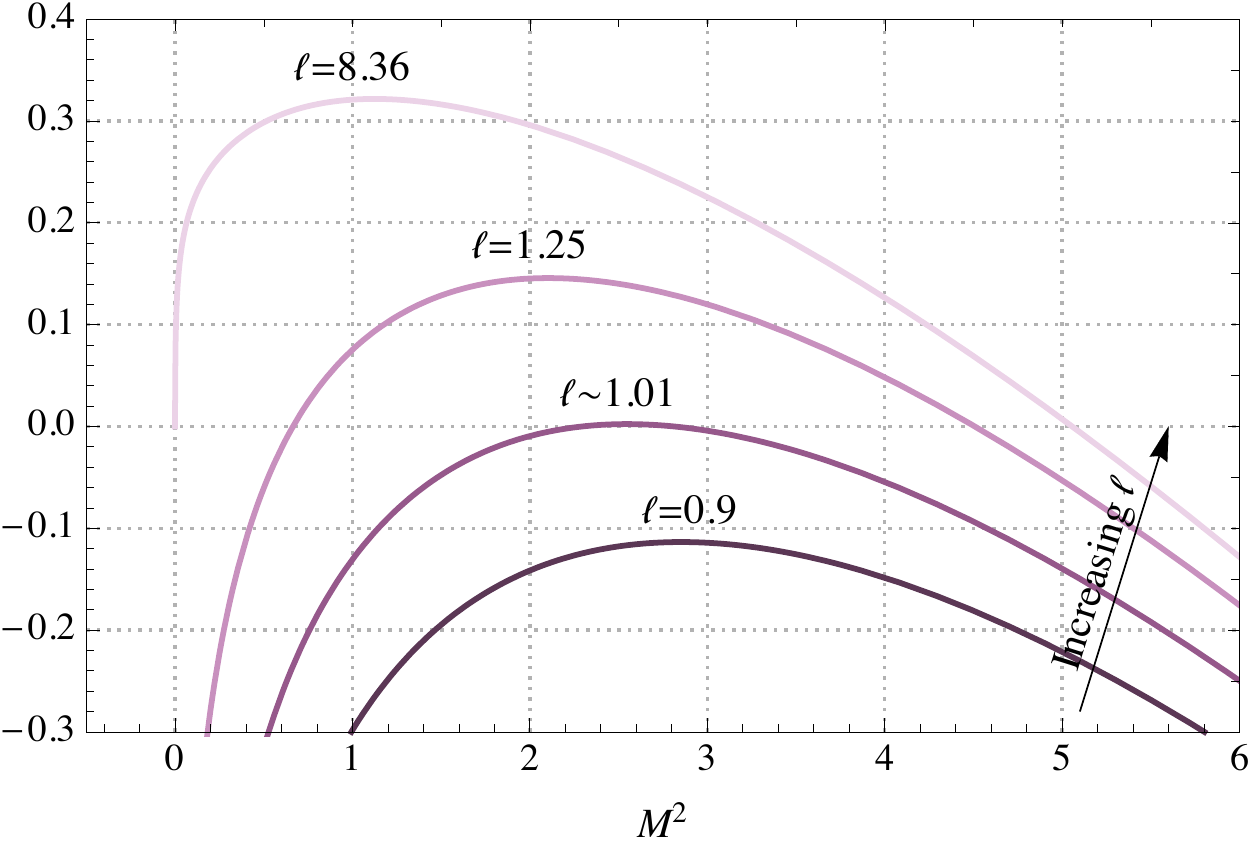}}
\hspace{0.3cm}
\subfigure{
\includegraphics[height=4.7cm,valign=t]{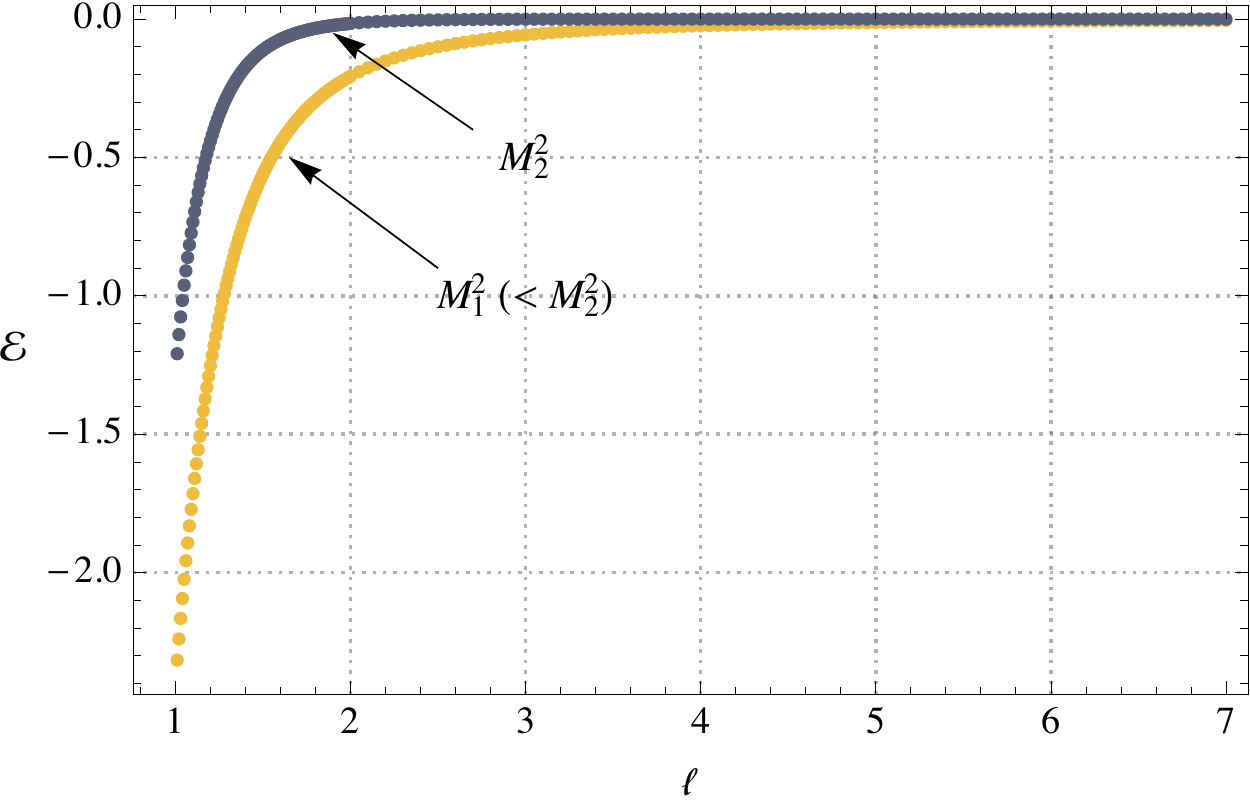}}
\end{array}$\\
\caption{{The number of roots of the gap equation depends on the competition between the three parameters, size, temperature and coupling constant. The {\em Left Panel} shows what happens varying the size of the system $\ell$ when temperature and coupling are fixed ($\beta=3$ and $\hat r=0.25$). There is no solution until $\ell\sim1.01$, which means no defined quantum vacuum. The theory becomes meaningful when roots appear, with the smaller of the two roots defining the ground state (compare with the {\em Right Panel}). At $\ell\sim 8.36$ the smaller of the two roots becomes zero, indicating the occurrence of symmetry breaking and the consequent appearance of Goldstone modes. The modular symmetry $\ell \leftrightarrow \beta$ ensures the same arguments to apply when temperature varies while $\ell$ and $\hat r$ are kept fixed.}}
\label{fig5}
\end{figure*}

At large $\beta$ and large $\ell$, the gap equation does not have a singular point in $M\to0$.
For large $\beta$ and small $\ell$, it is possible to follow the same arguments of section  \ref{secc} and expand $\varpi_3(0,\ell)$ for small size\footnote{In alternative, using \eqref{eq14} it is possible to show the general condition $\partial_{M^2} \varpi_D(\beta,\ell)=-\varpi_{D-2}(\beta,\ell)$},
\begin{align}
\varpi_3(0,\ell) &\approx -{16\pi^4\over 45}{1\over \ell^4} + \frac{4}{3} \pi^2 {M^2\over \ell^2}- \frac{8}{3} \pi {M^3\over \ell}-
M^4 \left(\gamma_e -\frac{3}{4}\right) - M^4 \log\left( {\ell M\over 4 \pi}\right) -{1\over 6}{\ell^2 M^6}\zeta'(-2).\label{final}
\end{align}
It is clear that $\partial_{M^2} \varpi_3(0,\ell)$ does not bring any singular contribution to \eqref{gapd3}, as one might expect. There are three different classes of solutions to the gap equation (see Fig.~\ref{fig5}). 
At fixed temperature and coupling constant, there is no solution for $\ell<\tilde{\ell}$, where $\tilde{\ell}$ depends on both $\beta$ and $\hat r$. This means that the quantum vacuum might be not well defined in the region, and the theory does not describe the system consistently. At $\ell=\tilde{\ell}$ a first (double) root appears, evolving in two different solutions as the size increases. The smaller of these two solutions turns out to be the vacuum ground state, the other being an excited state. At the critical length $\ell=\hat \ell$, the smaller root is $M^2=0$, signalling a phase transition in the coupling space: massless Goldstone modes appear, as the initial $\mathbb{O}(N)$ symmetry is broken into $\mathbb{O}(N-1)$. The same conclusions can be drawn if size $\ell$ is constrained to some value, and one let the temperature vary.  

\begin{wrapfigure}{r}{7cm}
\includegraphics[width=\linewidth]{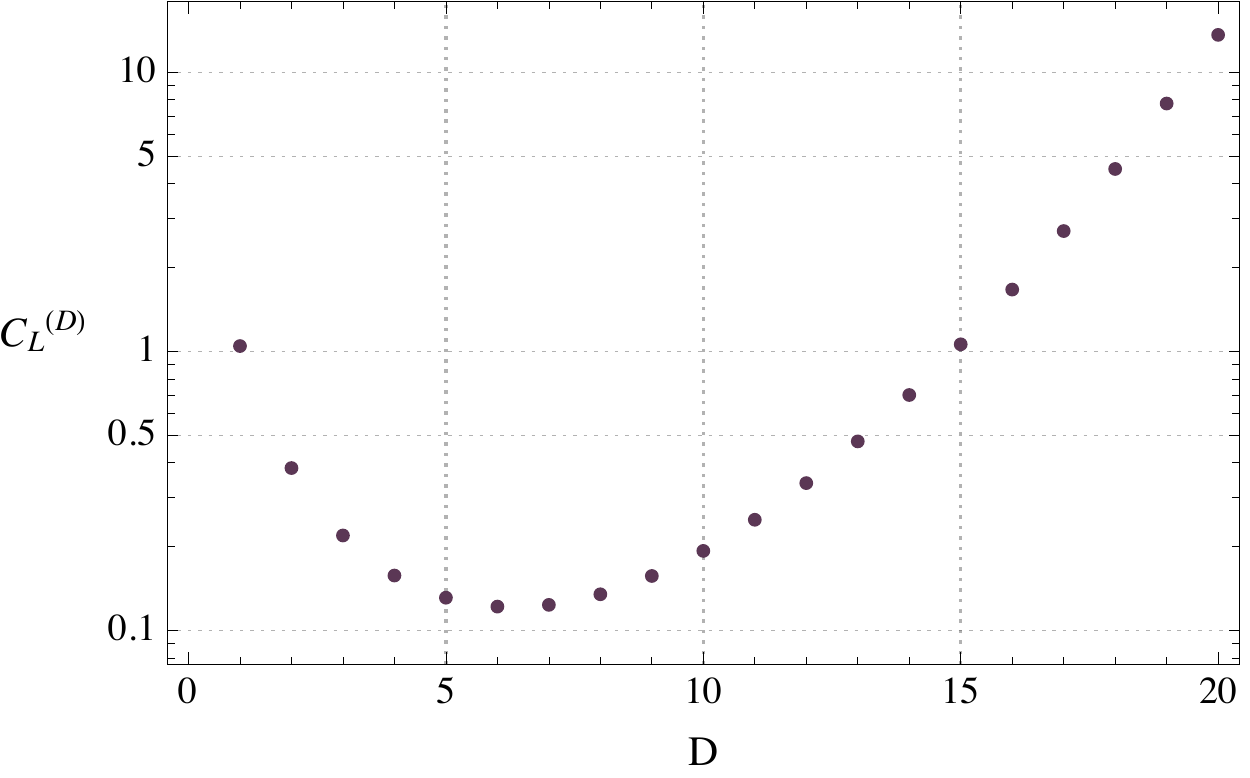}
{\vspace{-10pt} \caption{L\"uscher coefficients as a function of the dimension $D$. For $D=1$, it recovers the standard L\"uscher coefficient for a complex scalar field, $\mathcal{C}_L^{(1)} =\pi/3$. \vspace{1cm} \label{LuschD} }}
\end{wrapfigure}

The present calculation allows to isolate from the total free energy the coefficient of the  L\"uscher term, $-\mathcal{C}_L^{(D)}/L^D$, generalised to $D$ dimensions,
\begin{align}
\mathcal{C}_L^{(D)}=2~\pi^{-\frac{D+1}{2}}\cdot\Gamma\left(\frac{D+1}{2}\right)\cdot\zeta(D+1)\,,
\end{align}
whose values are plotted in Fig.~\ref{LuschD}. Interestingly, the coefficient with the minimum value is the one occurring at $D=6$.

\begin{figure*}[b!]
$\begin{array}{cc}
\subfigure{
\includegraphics[height=4.4cm,valign=t]{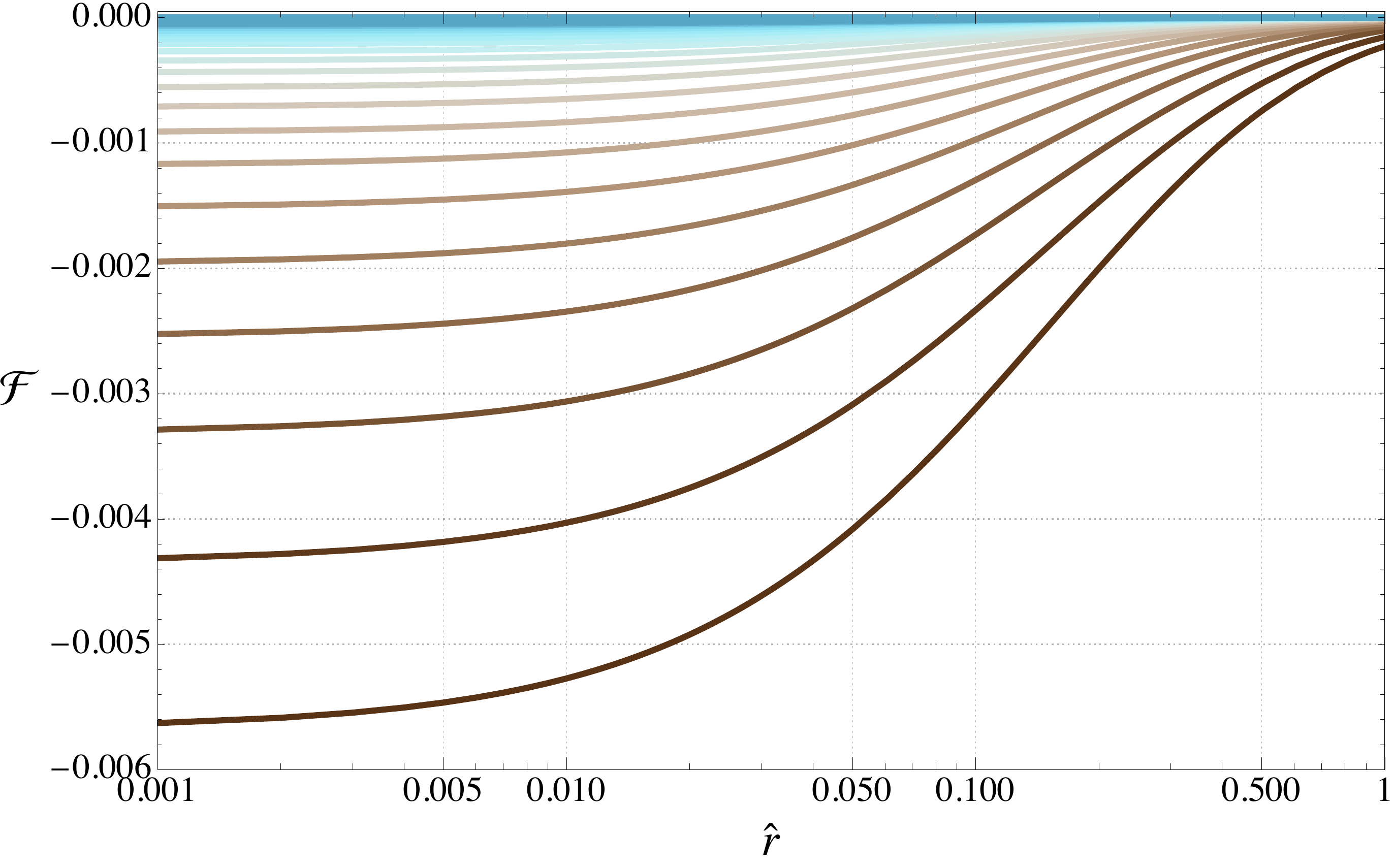}}
\subfigure{
\includegraphics[height=4.4cm,valign=t]{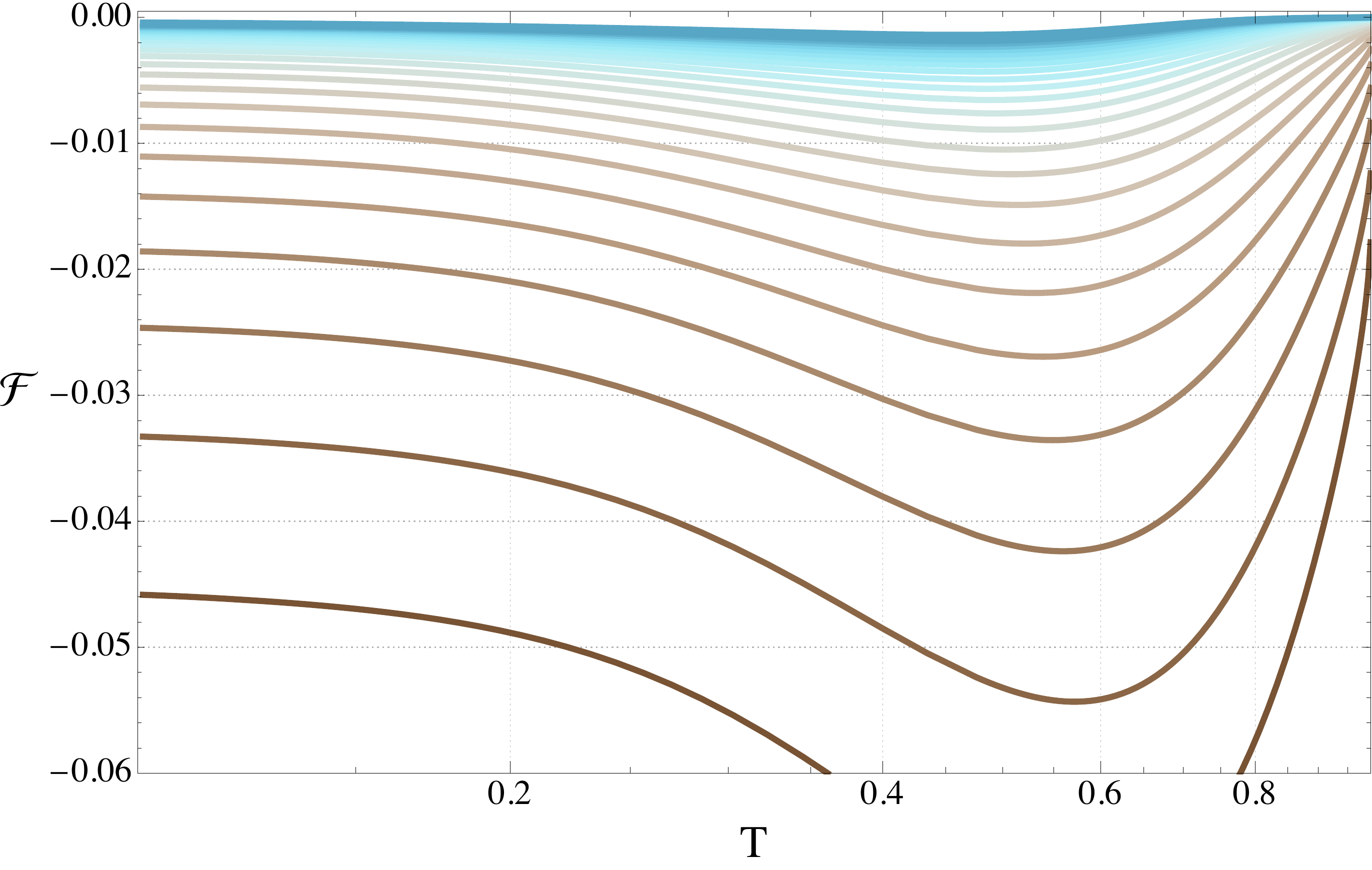}}
\hspace{0.16cm}
\subfigure{
\includegraphics[height=4cm,valign=t]{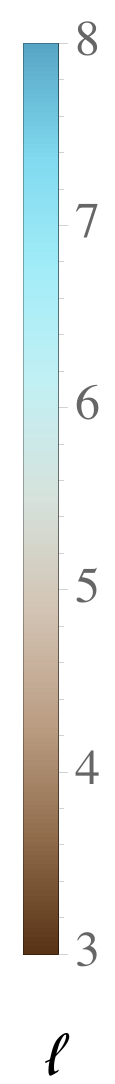}}
\end{array}$\\
\caption{{Casimir force in the allowed range of parameters for $D=3$ as a function of the rescaled coupling $\hat{r}$ ({\em Left Panel}, evaluated at $T=0.1$) and of the temperature $T$ ({\em Right Panel}, at $\hat{r}=0.25$). The color gradient corresponds to different sizes $\ell$, from $\ell=3$ (brown) to $\ell=8$ (cyan).}}
\label{fig6}
\end{figure*}

Fig.~\ref{fig6} summarises the resulting Casimir forces in the $D=3$ system. Again, for large $\hat r$, the mass gap is increasingly smaller. The Casimir energy is dominated by the higher-dimensional counterpart of the L\"uscher term (see first term of \eqref{final}), whose corresponding force is asymptotically approached by the curves in the {\em Left Panel} of Fig.~\ref{fig6}.  The modulation as a function of the temperature - {\em Right Panel} - is similar to the previous cases, with an important {\em caveat}: the picture is only valid until a certain critical temperature, depending on $\hat r$ and $\ell$, above which, as previously observed, the (formal) absence of a well-defined ground state suggests the loss of validity of the model.

\newpage

\section{Discussion and Conclusions}

While a nontrivial expression of the quantum vacuum arising in interacting field theories has been appreciated in various contexts (see, for example, Refs.~\cite{Flachi:2013bc,Flachi:2012pf,Flachi:2017cdo,Chernodub:2018pmt,Gambassi:2009,Gambassi2:2009,Gambassi3:2009} and references given there), the present set-up shows that substantial changes in the structure of the quantum vacuum occur even in simple quantum field theories and for the simplest choice of boundary conditions. Not only this leads to nontrivial modulation of the Casimir force as a function of external conditions ({\em e.g.}, couplings or temperature), but it also regulates the behaviour at small \textit{vs} large scales. This happens in the present case owing to the nonlinearities of the field theory that imply, in turn, a nonlinear dependence of the mass gap on the size of the system. Obviously the same is expected to generically happen in any interacting quantum field theory. 

This leaves us with the following important message: {\em substantial changes in Casimir forces can occur even for massive-nonlinear field theories, whereas one's expectation would be to see the forces exponentially decaying}. This statement is a direct consequence of the Mermin-Wagner-Hohenberg-Coleman theorem: quantum fluctuations in $D=1$ (and thermal fluctuations in $D=2$) forbid long-range interactions. The analytic regularization of the one-loop effective potential, indeed, unveils a logarithmic term proportional to $M^{1+D} \log M$ which, for $D=1$, prevents massless phases to take place (see also Refs. \cite{Bolognesi,Flachi:2019yci}). Imposing periodic boundary conditions, at any separation length $\ell$ the fluctuations acquire an effective mass. In terms of free (Casimir) energy, this dynamically generated mass induces an additional nonlinear dependence on the size of the system and, thereafter, a significant modulation of the Casimir force. 

An interesting follow-up of the present analysis concerns the prospect of a sign-flip in the force. It is known that the appropriate tuning of the boundary conditions allows for a change in the sign of the force (see for example Ref.~\cite{asorey:2013,Elizalde:2009nt}). Here, however, the boundary conditions are periodic, the simplest possible. A possible realisation of a system exhibiting a repulsive phase in the Casimir force even for periodic boundary conditions is the imperfect Bose gas in strongly anisotropic optical lattices \cite{Burgsmuller:2010nj,Jakubczyk1,Lebek:2020ppc}: whether this property could be effectively described through higher order operators in a ${\mathbb C}P^{N-1}$ or an $\mathbb{O}(N)$ model is currently under scrutiny.

The present findings should trigger further thinking about new ways to probe the quantum vacuum effects in quasi-one-dimensional cold-atomic systems, where nonlinear field theories describe relevant quantum fluctuations and boundary conditions can be mimicked by appropriate insertion of defects \cite{Jaksch:2005,Zohar:2016}. {The interest in the Casimir effect for these systems has also recently suggested the definition of a Casimir energy for lattice fermions \cite{Ishikawa:2020ezm}. Dirac matter quantum rings, quantum cylinders, and other lattice kirigamis with (anti-) periodic boundary conditions (now standard probes to study physical manifestations of the quantum vacuum, see e.g.  \cite{Castro:2018iqt,Flachi:2019btk,Cortijo:2011aa,deJuan:2010zz,Nissinen:2018dnq,Nissinen:2019kld}) are the possible arenas where consequences for the Casimir effect due to self-interactions among lattice fermions can be spotted.}

An interesting similarity is with colloidal particles immersed in binary liquid mixtures. While these systems are intrinsically higher-dimensional, once specific symmetries are appropriately imposed the dimensionality may be effectively lowered. Analogies with what we have discussed here with the Casimir effect in critical systems ({\em e.g.}, see Ref.~\cite{Gambassi:2009}) is certainly worth exploring.

\acknowledgments We acknowledge the support of the Japanese Society for the Promotion of Science (Grants-in-Aid for Scientific Research KAKENHI Grant n. 18K03626 and n. 17F17763) and of the Japanese Ministry of Education, Culture, Sports and Science (MEXT-supported Program for the Strategic Research Foundation at Private Universities `Topological Science' Grant No.\ S1511006). VV is supported by the H2020 programme and by the Secretary of Universities and Research of the Government of Catalonia through a Marie Sk{\l}odowska-Curie COFUND fellowship -- Beatriu de Pin{\'o}s programme no. 801370. VV thanks Keio University for hospitality during the initial stage of this work.

\end{document}